\documentclass[a4paper,12pt]{article}
\usepackage[margin=0.8in]{geometry}
\usepackage{graphicx}
\usepackage{epstopdf}
\usepackage{amssymb,amsmath}
\usepackage[colon,authoryear]{natbib}
\pdfoutput=1
\begin{document}

\title{On quasi-satellite periodic motion in asteroid and planetary dynamics}
\author{G. Voyatzis$^1$,  K. I. Antoniadou$^2$\vspace{0.25cm}\\
\footnotesize{$^1$Department of Physics, Aristotle University of Thessaloniki, 54124,
              Thessaloniki, Greece} \vspace{-0.15cm}\\
							\footnotesize{$^2$NaXys, Department of Mathematics, University of Namur, 8 Rempart de la Vierge, 5000 Namur, Belgium}\vspace{0.25cm}\\
              \footnotesize{voyatzis@auth.gr, kyriaki.antoniadou@unamur.be}     }
		
		\date{}  
\maketitle
\begin{center}
The final publication is available at\\ https://link.springer.com/article/10.1007/s10569-018-9856-2
\end{center}
\begin{abstract}
Applying the method of analytical continuation of periodic orbits, we study quasi-satellite motion in the framework of the three-body problem. In the simplest, yet not trivial model, namely the planar circular restricted problem, it is known that quasi-satellite motion is associated with a family of periodic solutions, called family $f$, which consists of 1:1 resonant retrograde orbits. In our study, we determine the critical orbits of family $f$ that are continued both in the elliptic and in the spatial model and compute the corresponding families that are generated and consist the backbone of the quasi-satellite regime in the restricted model. Then, we show the continuation of these families in the general three-body problem, we verify and explain previous computations and show the existence of a new family of spatial orbits. The linear stability of periodic orbits is also studied. Stable periodic orbits unravel regimes of regular motion in phase space where 1:1 resonant angles librate. Such regimes, which exist even for high eccentricities and inclinations, may consist dynamical regions where long-lived asteroids or co-orbital exoplanets can be found.      
\end{abstract}

{\bf keywords} 1:1 resonance -- co-orbital motion -- quasi-satellites -- periodic orbits -- three-body problem

\section{Introduction}
The term quasi-satellite (QS) motion refers to retrograde satellite motion, which takes place outside of the Hill's sphere \citep{Mikkola97}. In the framework of the three-body problem (TBP),  QS motion is a special case of 1:1 mean-motion resonance or, with another term, co-orbital motion. In the planar circular restricted TBP such a type of motion has been identified by the existence of the {\it family f} of periodic orbits \citep{Broucke68, Henon69, Benest74}. In the planar general TBP, where co-orbital planetary motion is considered, a family of periodic QS orbits has been computed by \cite{hpv09, hv11b}, and was called {\it family S}.       

Special interest for QS-orbits, which are called also {\it distant retrograde orbits} (DRO), is growing for the design of spacecraft missions around moons or asteroids \citep{Perozzi17, Minghu14}. A first application was achieved for the Phobos program \citep{SZ89} for which QS orbits were computed by \cite{Kogan89}. In the last twenty years, much attention has been given to QS asteroid motion. Evidence for the existence of stable QS-type motion around giant planets of our Solar system has been found after having considered analytical or semi-analytical perturbative methods \citep{Mikkola97, Namouni99, Nesvorny02, Mikkola06, Sidorenko14, Pousse17} or numerical integrations \citep{Wiegert00, Christou00a, Christou00b}. A lot of studies have  also focused on particular observed asteroids and Centaurs e.g. the 2002 VE68 around Venus \citep{Mikkola04} and the 2015 BZ509 around Jupiter \citep{Namouni18}. Furthermore, greatly interesting is the long-term stability of near-Earth asteroids, which are located in the QS regime, like e.g. 2004GU9, 2006FV35 and 2013LX28 \citep{Connors04, Morais06, Wajer10, Connors14} or of Earth-trojans \citep{Dvorak12}.

Exosolar planets in co-orbital motion, although having not been discovered yet, they can constitute exceptional planetary configurations, which include exomoons \citep{Heller18}, exotrojans or, possibly, quasi-satellite-like orbits, where planets of equivalent masses may evolve into \citep{Giuppone12, Funk13, Leleu17, Lillo18}. Studies on the possible existence and stability of such planetary configurations were performed for the co-planar case. For instance, a numerical study for the stability of exotrojans is given in \citet{Schwarz09}. \citet{hpv09} computed a stable family, $S$, of orbits which pass smoothly from planetary to satellite type orbits. In \citet{Giuppone10}, the stability regions are examined and, besides family $S$, new stable families, called anti-Lagrange solutions, are given by using a numerical averaging approach. In \citet{hv11b}, the existence of anti-Lagrange solutions is confirmed in the general TBP and it is shown that migration from planetary to satellite type of motion is possible under the effect of Stoke's-like dissipative forces. On the other hand, tidal forces may cause instabilities in co-orbital motion and planetary collisions \citep{Rodriguez13}. Analytical and semi-analytical treatment of the phase space structure of co-orbital motion has been given in \citet{Robutel13, Leleu17} by constructing appropriate averaged Hamiltonians for the 1:1 resonance.   

In an inertial frame of reference, QS motion is described by intersecting orbits of the small bodies and therefore, a resonant mechanism is necessary for avoiding close encounters \citep{Mikkola97, Mikkola06}. Studying the system in the framework of the TBP model in a rotating frame, periodic orbits are of major importance for understanding the underlying resonant dynamics. They consist exact positions of resonances and should appear as equilibrium points of a model, where the fast component of the motion is averaged. Linearly stable periodic solutions are associated with the existence of a foliation of invariant tori, which form a regime of long-term stability, where the resonant arguments librate regularly. For the 1:1 resonant QS motion, the main resonant argument is the angle $\theta=\lambda_2-\lambda_1$, where $\lambda_i$, $i=1,2$, denotes the mean longitude of the two bodies being in co-orbital motion. So, families of stable periodic orbits act as a guide for localising regimes which can host asteroids or planets in QS motion. 

In this study, by applying a methodological approach, we present a global view of the the main families of QS periodic orbits in all cases of the TBP model. By starting from the known family $f$ of the planar circular restricted TBP, we obtain bifurcations and compute the families of periodic orbits for more complicated versions of the TBP up to the general spatial TBP. The results for the planar models verify previous studies. Additionally, the three-dimensional motion is also studied, both in the restricted and in the general model. In Sect. \ref{res}, the restricted model is addressed, while in Sect. \ref{mass}, we apply continuation with respect to the mass of the smallest body and, thus, approach the families of the general TBP. In Sect. \ref{var}, we present the spatial families of the general model for various planetary masses and, finally, we conclude in Sect. \ref{con}.     

\section{QS periodic motion in the restricted three-body model (RTBP)}\label{res}
In the restricted case, we consider the system Sun - Planet - asteroid, where the Sun and the Planet (primary bodies) have masses $m_0$ and $m_1$, respectively, and revolve around their centre of mass $O$ in a Keplerian orbit.  We consider the classical rotating frame $Oxyz$, where the $Oxy$ plane coincides with the inertial one that contains the orbit of the primaries, the $x$-axis is directed along the direction line Sun - Planet and $Oz$ is perpendicular to the plane $Oxy$. In this frame, and by considering the mass normalisation $m_0+m_1=1$, the mass parameter $\mu=m_1$ and the gravitational constant $G=1$, the motion of the massless asteroid (body 2) is described by the Lagrangian function
\begin{equation}\label{LagrangeR}
\mathcal{L_R}=\mathcal{T_R}-\mathcal{U_R},
\end{equation}
where
$$
\begin{array}{l}
\mathcal{T_R}=\frac{1}{2} \left ( \dot x^2+\dot y^2+\dot z^2 \right ) + \left (x\dot y-y\dot x \right )\dot \upsilon + \frac{1}{2} \left ( x^2+y^2 \right ) \dot \upsilon^2, \vspace{0.2cm}\\
\mathcal{U_R}=\displaystyle{-\frac{1-\mu}{r_{02}}-\frac{\mu}{r_{12}}}, \vspace{0.2cm} \\
r_{02}=\sqrt{(x+\mu r_{01})^2+y^2+z^2},\quad r_{12}=\sqrt{(x-(1-\mu) r_{01})^2+y^2+z^2},
\end{array}
$$
and $\upsilon=\upsilon(t)$ is the true anomaly and $r_{01}=r_{01}(t)$ the mutual distance of the primaries along their relative Keplerian orbit. 

In the following, we will refer also to the planetocentric, barycentric and heliocentric osculating orbital elements of the orbits, $a_i$ (semi-major axis), $e_i$ (eccentricity), $\varpi_i$ (longitude of pericenter), $\Omega_i$ (longitude of ascending node) and $\lambda_i$ (mean longitude), where the index $i=1$ and 2 refers to the planet and the asteroid, respectively.  

\subsection{The planar circular restricted model (PC-RTBP)}
\begin{figure}
\centering
\includegraphics[width=1 \textwidth]{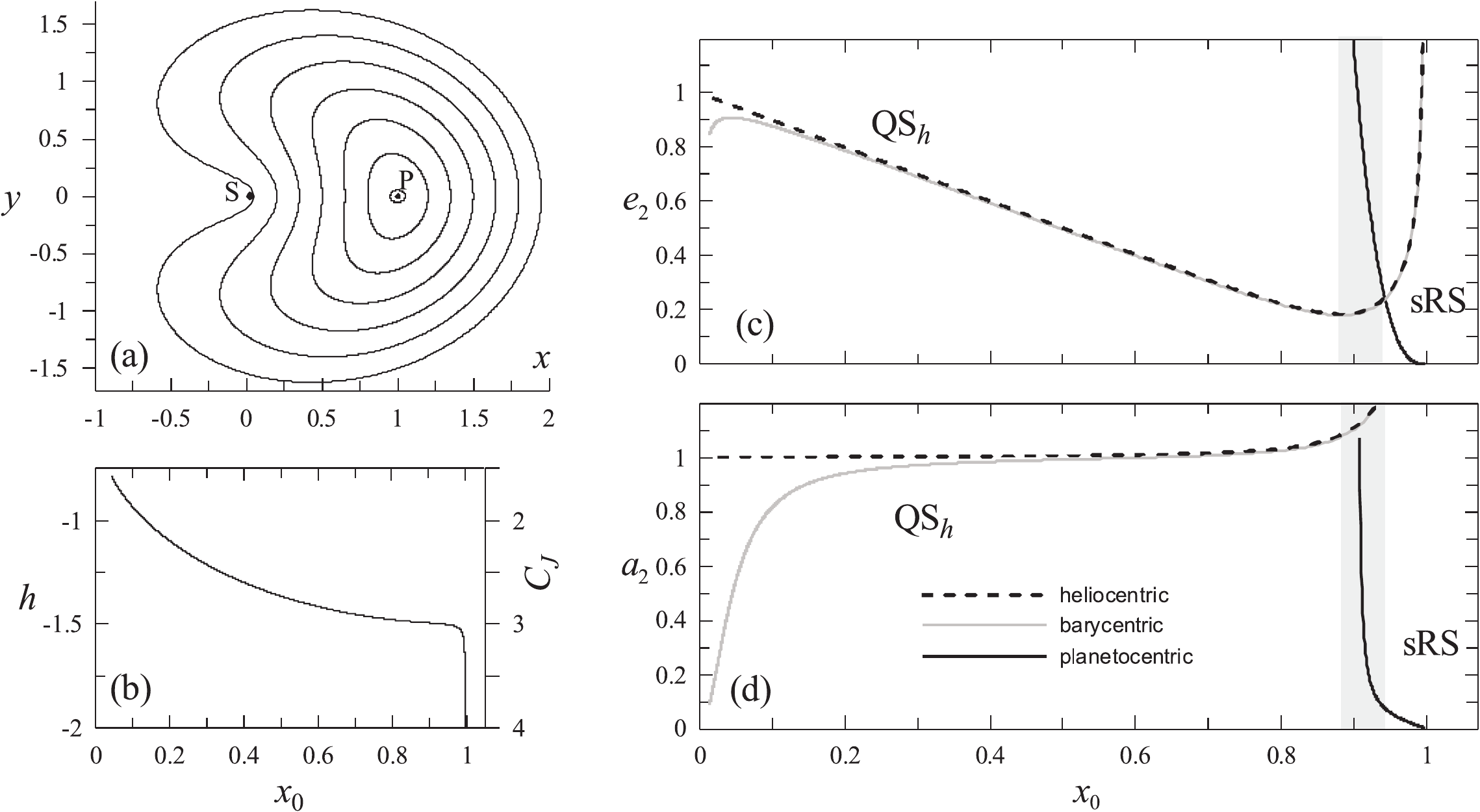}
\caption{The family $f$ of periodic orbits for $\mu=0.001$. {\bf a} Orbits in the rotating frame, where $S$ and $P$ indicate the Sun and the Planet, respectively. {\bf b} The characteristic curve $x_0 - h$ of the family $f$. {\bf c} The eccentricity and {\bf d} the semi-major axis of the orbits along family $f$ computed for the heliocentric, barycentric and planetocentric reference system. The grey zone  indicates the $\textnormal{QS}_b$ domain defined in \citet{Pousse17}, which separates the heliocentric quasi-satellite ($\textnormal{QS}_h$) orbits from  retrograde satellite orbits (sRS)} 
\label{FIG_FAMF}
\end{figure}

Considering the primaries moving in a circular orbit with unit mutual distance ($e_1=0$, $a_1=1$) and the asteroid on the plane $Oxy$ ($z=0$) we obtain the planar circular restricted three-body problem, where the Sun and the Planet are fixed on the $x$-axis at position $-\mu$ and $1-\mu$, respectively \citep{murraydermott}. We have $r_{01}=1$ and $\dot\upsilon=1$, thus, the system (\ref{LagrangeR}) is autonomous of two degrees of freedom and possesses the Jacobi integral $C_J$ or the equivalent ``energy integral" $h$ given by 
\begin{equation}\label{energyRC}
h= \frac{1}{2} \left (\dot x^2+\dot y^2 \right ) - \frac{1}{2} \left (x^2 + y^2 \right ) - \frac{1-\mu}{\sqrt{(x+\mu)^2+y^2}}-\frac{\mu}{\sqrt{(x-1+\mu)^2+y^2}}=-C_J/2.
\end{equation}  
In this model, QS orbits are associated with the existence of a family of symmetric periodic orbits, called family $f$ \citep{Broucke68, HenonGuyot70, Pousse17}. This family tends to the H\'enon's family $E_{11}^+$ of generating orbits which starts from a third-species orbit, as $\mu\rightarrow 0$ (i.e. the orbit of the asteroid coincides with the planet's) and terminates at a collision orbit with the Sun \citep{henon97}. Thus, at least for small values of $\mu$, family $f$ starts with orbits that encircle the planet at average distance that approaches zero. As the distance from the planet increases, the family terminates at a collision orbit with the Sun. For $\mu=0.001$, periodic orbits along family $f$ are shown in the rotating frame in Fig. \ref{FIG_FAMF}a. They can be assigned to initial conditions $x_0$, $y_0=\dot x_0=0$ and $\dot y_0$, where, from Eq. \eqref{energyRC}, $\dot y_0=\dot y_0(x_0,h)$. Assuming the interval $-\mu < x_0 <1-\mu$, the orbits of family $f$ can be mapped to a unique value $x_0$, which can be used as the parameter of the family. The characteristic curve $x_0 - h$ of the family is shown in Fig. \ref{FIG_FAMF}b. 

When the motion of the massless body takes place close to the Planet, where the gravitational perturbation of the Sun is assumed relatively very small, we obtain almost Keplerian retrograde satellite orbits. When the orbits of the family $f$ are quite distant from the Planet, the gravitation of the Sun dominates and the orbits are almost Keplerian planetary-type orbits. So, we can describe the orbits by using osculating orbital elements and by computing them in a planetocentric system (for satellite orbits) or in a heliocentric system (for planetary type orbits). The distinction between the two types of orbits cannot be strictly defined. In panels (c) and (d) of Fig. \ref{FIG_FAMF}, we present the eccentricity, $e_2$, and the semi-major axis, $a_2$, for the orbits of family $f$ considering a heliocentric, a barycentric and a planetocentric reference system. For all orbits the longitude of perihelion is $\varpi_2=0^\circ$. In the plots, we present also the regions sRS (retrograde satellite orbits), $\textnormal{QS}_h$ (heliocentric quasi-satellite orbits) and the separation grey region (called $\textnormal{QS}_b$, binary quasi-satellite), which are defined in \citet{Pousse17}. The right border of the grey region, measured from the planet position, is the Hill's radius $R_H$, which can alternatively be used for the distinction between $\textnormal{QS}_h$ and sRS orbits. We can observe that as sRS orbits approach the Planet ($x_0\rightarrow 1-\mu$) their planetocentric eccentricity tends to zero. The heliocentric eccentricity of $\textnormal{QS}_h$ orbits increases and tends to 1, as we approach the collision orbit with the Sun. The slope of the increasing eccentricity, as $x_0$ decreases, is almost equal to 1, because $x_0$ is the perihelion distance and $a_2\approx 1$, as it is shown in panel (d). Apparently, the periodic orbits of the family $f$ indicate the exact position of the 1:1 mean motion resonance of QS orbits \citep{Sidorenko14}.    

\begin{figure}
\centering
\includegraphics[width=1 \textwidth]{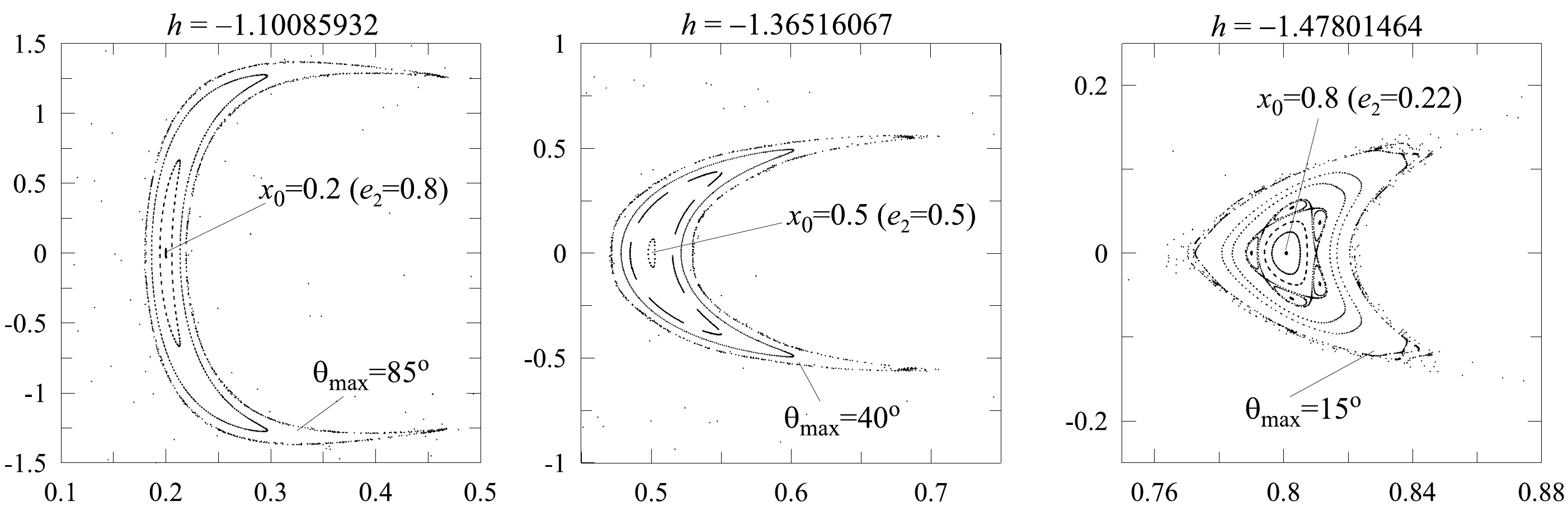}
\caption{Poincar\'e sections on the plane $x-\dot x$ ($y=0$, $\dot y>0$) for different energy levels. For each case the initial position $x_0$ and the corresponding eccentricity of the QS periodic orbit is indicated. The maximum amplitude of librations of the resonant angle $\theta$, which is computed for  the last invariant curve (approximately) of the island of stability, is indicated. Strong chaos occurs outside of the islands} 
\label{FIG_PNC}
\end{figure}             

All periodic orbits of family $f$ are linearly (horizontally) stable for $\mu<0.0477$ \citep{Benest74}. Consequently, in a Poincar\'e surface of section they are presented as fixed points surrounded by invariant tori that form ``islands of stability''. Using the rotating reference frame, we present in Fig. \ref{FIG_PNC} surfaces of section $y=0$ ($\dot y>0$) for some energy values in the $\textnormal{QS}_h$ regime. Outside the islands of stability strongly chaotic motion occurs. For the regular orbits the resonant angle $\theta=\lambda_2-\lambda_1$ librates. The maximum amplitude of libration, $\theta_{max}$, which corresponds to the orbit of the last invariant tori of the island region, increases as the orbits become more distant from the Planet. This has been shown also by \citet{Pousse17} with the use of an average model. However, due to the absence of chaos in the averaged model, the maximum amplitude of libration is overestimated.     

\subsection{The planar elliptic model (PE-RTBP)} \label{PERTBP}
Assuming the primaries moving in an eccentric orbit ($e_1\neq 0$, $a_1=1$), the system \eqref{LagrangeR} becomes non-autonomous and periodic in time with period $T'$ equal to the period of the revolutions of the primaries, namely in our units $T'=2\pi$. Concerning the continuation of periodic orbits from the circular to the elliptic model, this is possible for the periodic orbits of the circular model, which have period $T_c=\frac{p}{q}T'$, with $p$ and $q$ being prime integers. Starting from such a periodic orbit, which corresponds to $e_1=0$, we can obtain by analytic continuation monoparametric families of periodic orbits for $e_1\neq 0$. Along these families, the period of orbits is constant, $T=q\,T_c=2p\pi$, and $q$ defines the multiplicity of the orbits \citep{Broucke69}. Particularly, two distinct families are generated according to the initial location of the primary (perihelion or aphelion).

\begin{figure}
\centering
\includegraphics[width=1 \textwidth]{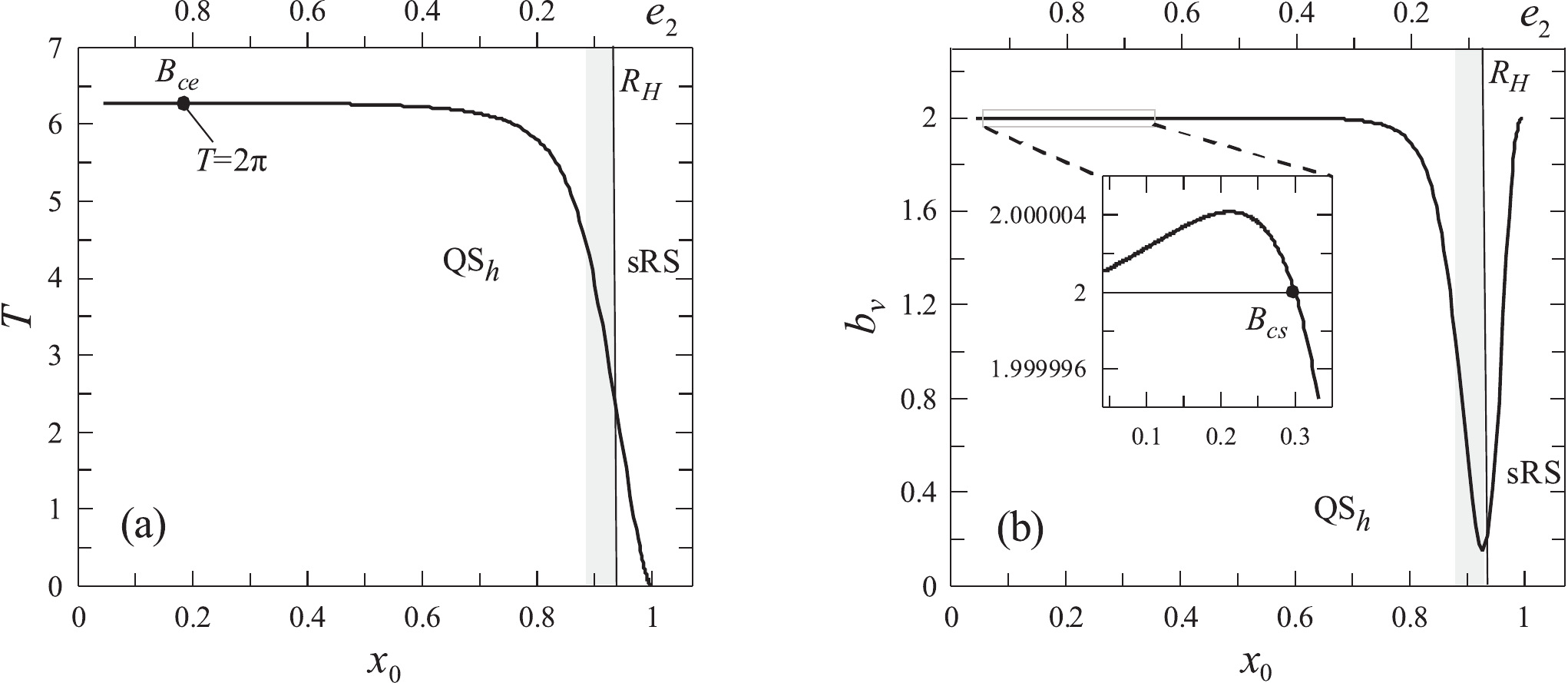}
\caption{{\bf a} The period $T$ of orbits along the family $f$.  {\bf b} The vertical stability index along the family $f$. The vertical line indicates the location of Hill's radius, $R_H$, which is the right border of the $\textrm{QS}_b$ region (gray zone).} 
\label{FIG_TVS}
\end{figure}

In Fig. \ref{FIG_TVS}a, we present the variation of the period along the family $f$ of the circular model. We have added an axis showing the eccentricity $e_2$, which can be used also as the parameter of the family $f$ in the $\textnormal{QS}_h$ regime. Considering the case of simple periodic orbits ($q=1$), we obtain  the periodic orbit $B_{ce}$ of period $T=2\pi$, eccentricity $e_2=0.8356$ and semi-major axis $a_2=1.0014$ along the family $f$. This orbit can be assumed as a bifurcation (or generating) orbit for a family of periodic QS orbits in the PE-RTBP.  Certainly, many other cases of higher multiplicity can be obtained in $\textnormal{QS}_h$ regime.  \citet{Lidov94} and \citet{vgv12} used the PE-RTBP and the elliptic Hill model, respectively, and studied multiple QS periodic orbits ($q\geq 2$).      

\begin{figure}
\centering
\includegraphics[width=1 \textwidth]{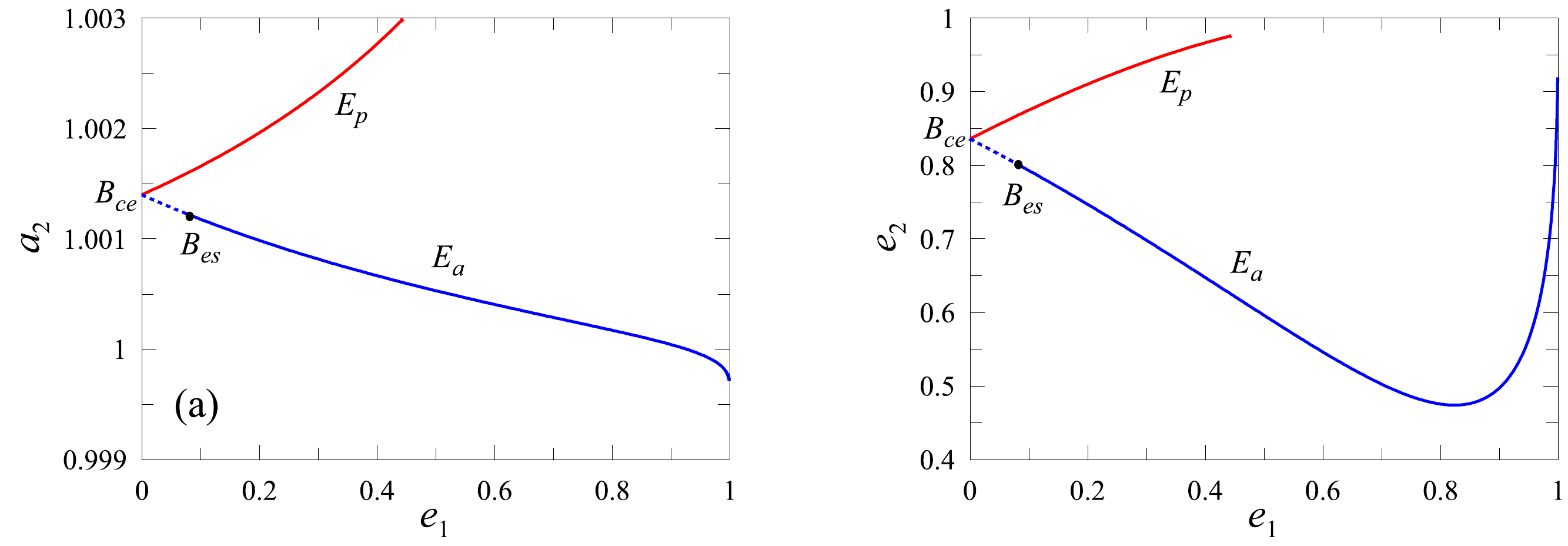}
\caption{The families $E_p$ and $E_a$ of the PE-RTBP of simple multiplicity ($T=2\pi$). Characteristic curves with parameter the eccentricity of the primaries ($e_1$) in the plane {\bf a}  $e_1 - a_2$  and {\bf b}  $e_1 - e_2$. Blue solid segments indicate horizontal and vertical stability, dotted parts are horizontally stable but vertically unstable and red parts are both horizontally and vertically unstable} 
\label{FIG_Efams}
\end{figure} 

In Fig. \ref{FIG_Efams}, we present the families, $E_p$ (for $\varpi_1=0$, $\upsilon(0)=0$) and $E_a$ (for $\varpi_1=\pi$, $\upsilon(0)=\pi$), which bifurcate from $B_{ce}$. All orbits of family $E_p$ are horizontally and vertically unstable. The continuation of the family becomes computationally very slow for $e_1>0.44$, where $e_2\rightarrow 1$. Family $E_a$ continues up to $e_1=1$ (rectilinear model, see \cite{vtg18}) and along it the asteroid's eccentricity, $e_2$, initially decreases. At $e_1=0.825$ it takes its minimum value ($\sim 0.474$) and increases afterwards. It consists of horizontally stable periodic orbits, but as $e_2\rightarrow 1$, they seem to become unstable. Since $B_{ce}$ is vertically unstable (see section \ref{SCRTBP}), family $E_a$ also starts with vertically unstable periodic orbits, but at $e_1=0.0858$ (orbit denoted by $B_{es}$) the orbits turn into vertically stable. So, $B_{es}$ is a vertical critical orbit and is a potential generating orbit for a family of the spatial model, as we will see in Sect. \ref{SERTBP}. 

We mention that the families $E_a$ and $E_p$ have been also computed by \citet{Pousse17}, as sets of equilibrium solutions (called $G_{QS,1}^{e'}$ and $G_{QS,2}^{e'}$, respectively)  by using a numerically averaged model. Their results are in a very good agreement with those presented in Fig. \ref{FIG_Efams}. A worthy noted difference is that $G_{QS,1}^{e'}$ is stable up to $e_1\approx 0.8$, while the equivalent family $E_a$ is stable up to $e_1\rightarrow 1$. Also, the perturbative approach used by \citet{Mikkola06} showed the stability of QS planar orbits under the perturbation caused by the elliptic orbit of the primaries and by assuming small inclinations. However, it could not provide the above periodic solutions, due to the high eccentricities.       

\subsection{The spatial circular problem (SC-RTBP)} \label{SCRTBP}

Vertical stability and three-dimensional families emanating from the short and long period planar families of co-orbital trojan-like orbits have been studied extensively (see e.g. \cite{Perdios91,Hou08}). Here, we consider QS co-orbital motion and examine the planar orbits of family $f$ with respect to their vertical stability. For each orbit of period $T$ we compute the index 
\begin{equation}\label{Vindex}
b_v=|\textnormal{trace} \Delta(T)|,
\end{equation}
where $\Delta(T)$ is the monodromy matrix of the vertical variations \citep{henon73}. The orbit is vertically stable iff $|b_v|<2$. Orbits with $|b_v|=2$ are called vertical critical orbits (v.c.o.) and they can be analytically continued to the spatial problem, for $z(0)\neq 0$ or $\dot{z}(0)\neq0$. 

In Fig. \ref{FIG_TVS}b, we present the index $b_v$ along the family $f$. The index $b_v$ exhibits a minimum at the orbit located at the Hill's radius. Then, in the sRS regime, the index increases, but $b_v<2$ always holds. In the $\textnormal{QS}_h$ regime, $b_v$ is close to the critical value 2 for a long interval, but actually exceeds the critical value at the orbit $B_{cs}$, where $x_0\approx 0.3$ and $e_2\approx 0.7$. 

\begin{figure}
\centering
\includegraphics[width=1 \textwidth]{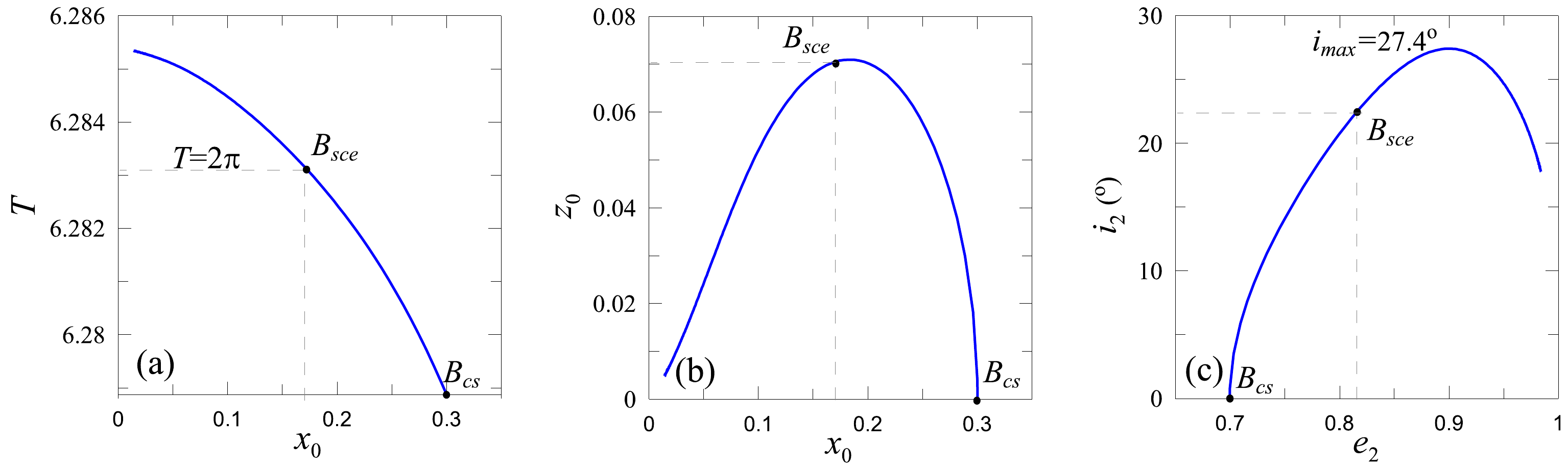}
\caption{Projections of the spatial family $F$ on the planes {\bf a} $x_0-T$, {\bf b} $x_0-z_0$ and {\bf c} $e_2-i_2$. The family starts from the v.c.o. $B_{cs}$. The orbit $B_{sce}$ is the orbit of $F$ with period $T=2\pi$} 
\label{FIG_Ffam}
\end{figure} 

For the v.c.o. $B_{cs}$, we can apply numerical continuation by using differential corrections and obtain a family of spatial periodic orbits in the spatial circular model ($e_1=0$, $z\neq 0$). We call this family $F$ with its orbits being identified by the non-zero initial conditions ($x_0$, $\dot{y}_0$, $z_0$) beside $y_0$=$\dot{x}_0$=$\dot{z}_0$=0. In Fig. \ref{FIG_Ffam}a, we present the evolution of the period $T$ along the family (using $x_0$ as parameter). $T$ increases monotonically and takes the value $2\pi$ at $x_0\approx 0.16$. This orbit, denoted by $B_{sce}$, is potential for continuation in the spatial elliptic model. In panel (b), we obtain that $z_0$ takes a maximum value for $x_0=0.183$ and then decreases towards zero as $x_0\rightarrow 0$ (close approach to the Sun). The characteristic curve in the eccentricity - inclination plane is shown in panel (c). The whole family is located in the high eccentricity regime and the maximum inclination observed is $\sim 27^\circ$. The orbit $B_{sce}$ has inclination $22.5^\circ$.  All orbits of the family $F$ are linearly stable with $\varpi_2=0$.

\subsection{The spatial elliptic problem (SE-RTBP)} \label{SERTBP}
We compute the index given by Eq. (\ref{Vindex}) for the elliptic planar model too, and particularly, for the orbits of the families $E_p$ and $E_a$. As we can see from Fig. \ref{FIG_TVS}b, the orbit $B_{ce}$, where these families originate, is vertically unstable. Thus, both families start with vertically unstable orbits. As we mentioned in Sect. \ref{PERTBP}, $E_p$ is whole vertically unstable, while $E_a$ becomes vertically stable after the v.c.o. $B_{es}$, which can be analytically continued  to the spatial problem providing a family of spatial periodic orbits \citep{ichmich80}. Also, as we mentioned in Sect. \ref{SCRTBP}, the spatial orbit $B_{sce}$ is also a potential orbit for a continuation with $e_1\neq 0$ and with the Planet located initially at its perihelion or aphelion. 

\begin{figure}
\centering
\includegraphics[width=0.6 \textwidth]{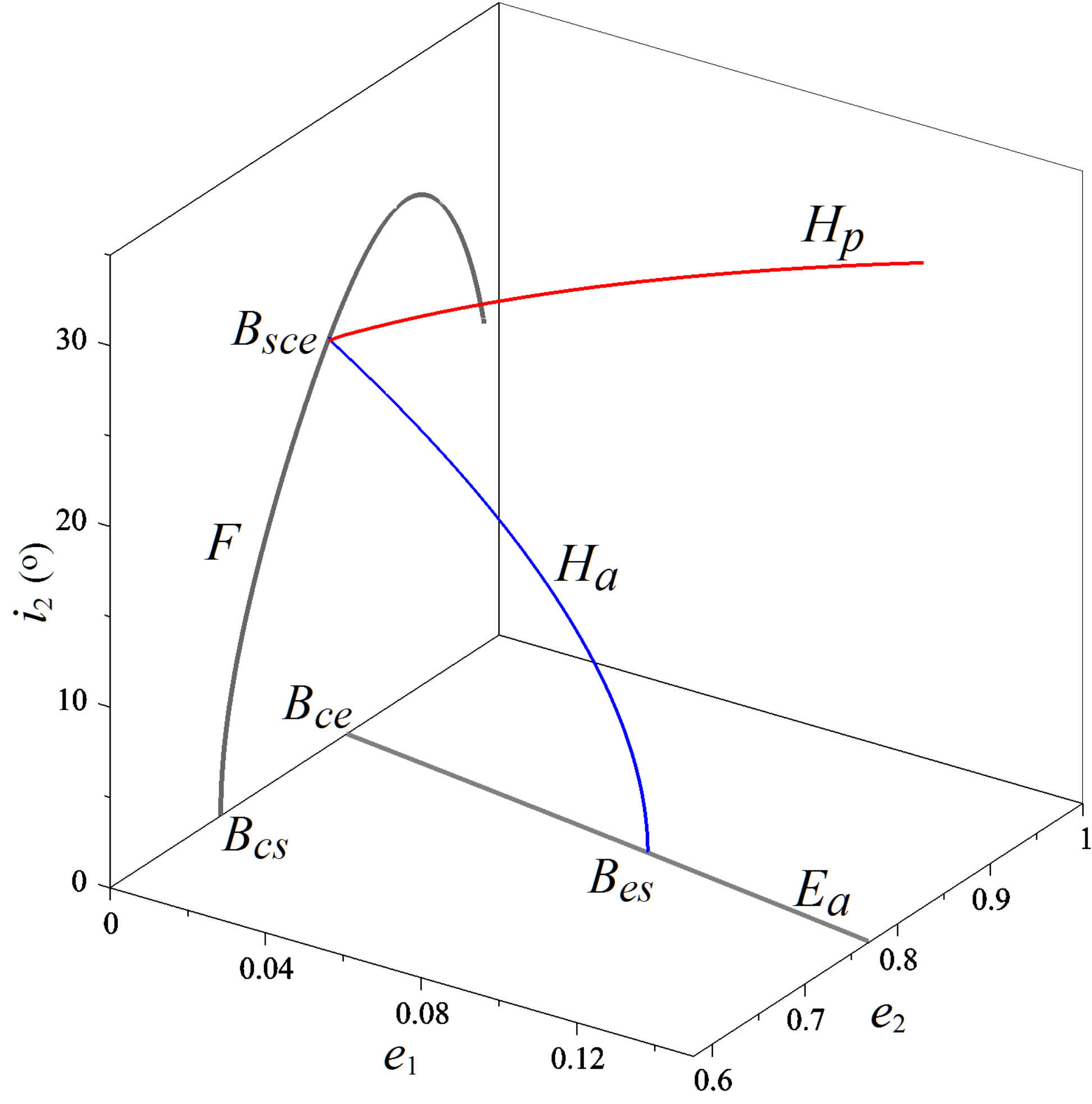}
\caption{The families $H_p$ and $H_a$ of the SE-RTBP on the projection space $e_1-e_2-i_2$. Along the families $a_2\approx 1.0$ ($a_1=1$). Family $H_p$ is linearly unstable, while $H_a$ is stable} 
\label{FIG_Hfams}
\end{figure} 

By considering the planet at the perihelion, the continuation of the orbit $B_{sce}$ for $e_1\neq 0$ provides the family $H_p$. This family is linearly unstable and extends up to very high eccentricities, $e_1\approx 0.91$ and $e_2\rightarrow 1$ and high inclination value, $i_2\approx 58^\circ$. The initial segment of $H_p$ is presented in Fig. \ref{FIG_Hfams}. 
   
By continuing the periodic orbits $B_{es}$ for $i_2\neq 0$ and $B_{sce}$ for $e_1\neq 0$ (with the Planet at aphelion at $t=0$), we obtain a unique stable family, denoted by $H_a$ (Fig. \ref{FIG_Hfams}). Thus, family $H_a$ forms a \textit{bridge} linking the families $E_a$ and $F$. The inclination along the family monotonically increases from $B_{es}$ ($i_2=0^\circ$) to $B_{sce}$ ($i_2=22.5^\circ$).

\section{From the restricted to the general three-body problem (GTBP)}\label{mass}

We consider the general three-body problem in a configuration ``Sun - Planet - third body'',  where the third body is a second planet or a satellite. We will use the indices 0, 1 and 2 for referring to the three bodies, respectively. In the GTBP, it is $m_2\neq 0$ and, based on an inertial frame $OXYZ$, with $O$ being the centre of mass, we can assume a rotating frame $Gxyz$ where i) the origin $G$ is the centre of mass of $m_0$ and $m_1$  ii) $Gz$-axis is parallel to $OZ$ and iii) the bodies $m_0$ and $m_1$ move always on the plane $Gxz$ \citep{mich79}. We denote by $\upsilon$ the angle between the axes $Gx$ and $OX$. In this rotating frame, the Lagrangian is written 
\begin{equation}\label{LagrangeG}
\mathcal{L_G}=\frac{1}{2} M_1\left (\dot x_1^2+\dot z_1^2+x_1^2 \dot \upsilon^2 \right )+M_2\mathcal{T_R}-\mathcal{U_G},
\end{equation}
where in $\mathcal{T_R}$ the coordinates $x$,$y$ are denoted now by $x_2$, $y_2$ and
$$
\begin{array}{l}
M_1=\frac{(m_0+m_1)m_1}{m_0}, \;M_2=\frac{(m_0+m_1)m_2}{m_0+m_1+m_2} ,\;\; 
\mathcal{U_G}=\displaystyle{-\sum_{i,j=0}^2 \frac{m_i m_j}{r_{ij}} \;\;(i\neq j)}, \vspace{0.2cm} \\
r_{01}^2=(1+a)^2 (x_1^2+z_1^2),\; a=\frac{m_1}{m_0},  \vspace{0.2cm}\\
r_{02}^2= (a x_1+x_2)^2+y_2^2+(az_1+z_2)^2, \;\quad r_{12}^2=(x_1-x_2)^2+y_2^2+(z_1-z_2)^2,
\end{array}
$$
We use the mass normalisation $m_0+m_1+m_2=1$ and note that for $m_2\rightarrow 0$ the Lagrangian (\ref{LagrangeG}) provides the same equations of motion with those of (\ref{LagrangeR}). Apart from the Jacobi integral the vector of angular momentum $\mathbf{L}=(0,0,p_\upsilon)$, where $p_\upsilon=\partial \mathcal{L_G} / \partial{\dot\upsilon}$, is also conserved and provides $z_1$, $\dot z_1$ and $\dot\upsilon$ as functions of the variables ($x_1$, $x_2$, $y_2$ and $z_2$) and their time derivatives \citep{mich79, kat86, av13b}.  In the following, we will present the characteristic curves with respect to their osculating orbital elements that correspond to the initial conditions. For symmetric periodic orbits the initial angles $\Delta \varpi=\varpi_2-\varpi_1$, $\Delta\Omega=\Omega_2-\Omega_1$ and $\theta=\lambda_2-\lambda_1$ are either equal to $0$ or $\pi$. Also, we will refer to the mutual inclination of the small bodies, $\Delta i$. 

\subsection{The planar general problem (P-GTBP)}
According to \cite{Hadjidem75}, all periodic orbits of the PC-RTBP (where $m_2=0$) can be continued for $0<m_2\ll 1$ with the same period, $T$, provided that their period is not an integer multiple of the period of the primaries, i.e. $T\neq 2k\pi$, $k \in N$, in our normalisation. A direct deduction is that all QS periodic orbits of family $f$ are continued to the P-GTBP except the orbit $B_{ce}$, which is the generating orbit of the families $E_a$ and $E_p$ in the PE-RTBP. Also, all periodic orbits of the PE-RTBP ($e_1\neq 0$) are continued to the P-GTBP, but with different periods \citep{ichkm78, avk11}. 

\begin{figure}
$
\begin{array}{ccc}
\includegraphics[width=0.45\textwidth]{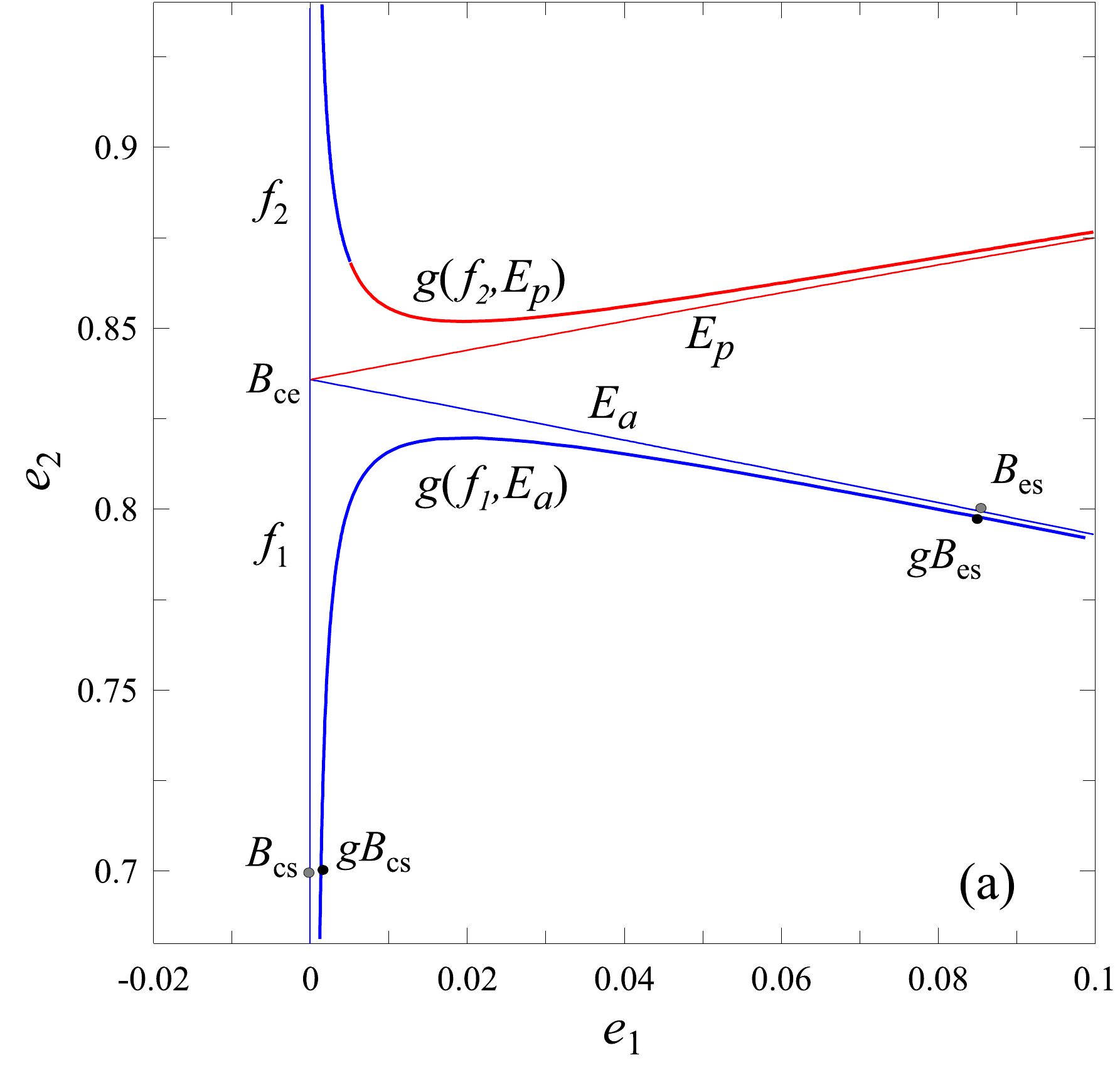} & \quad & \includegraphics[width=0.45\textwidth]{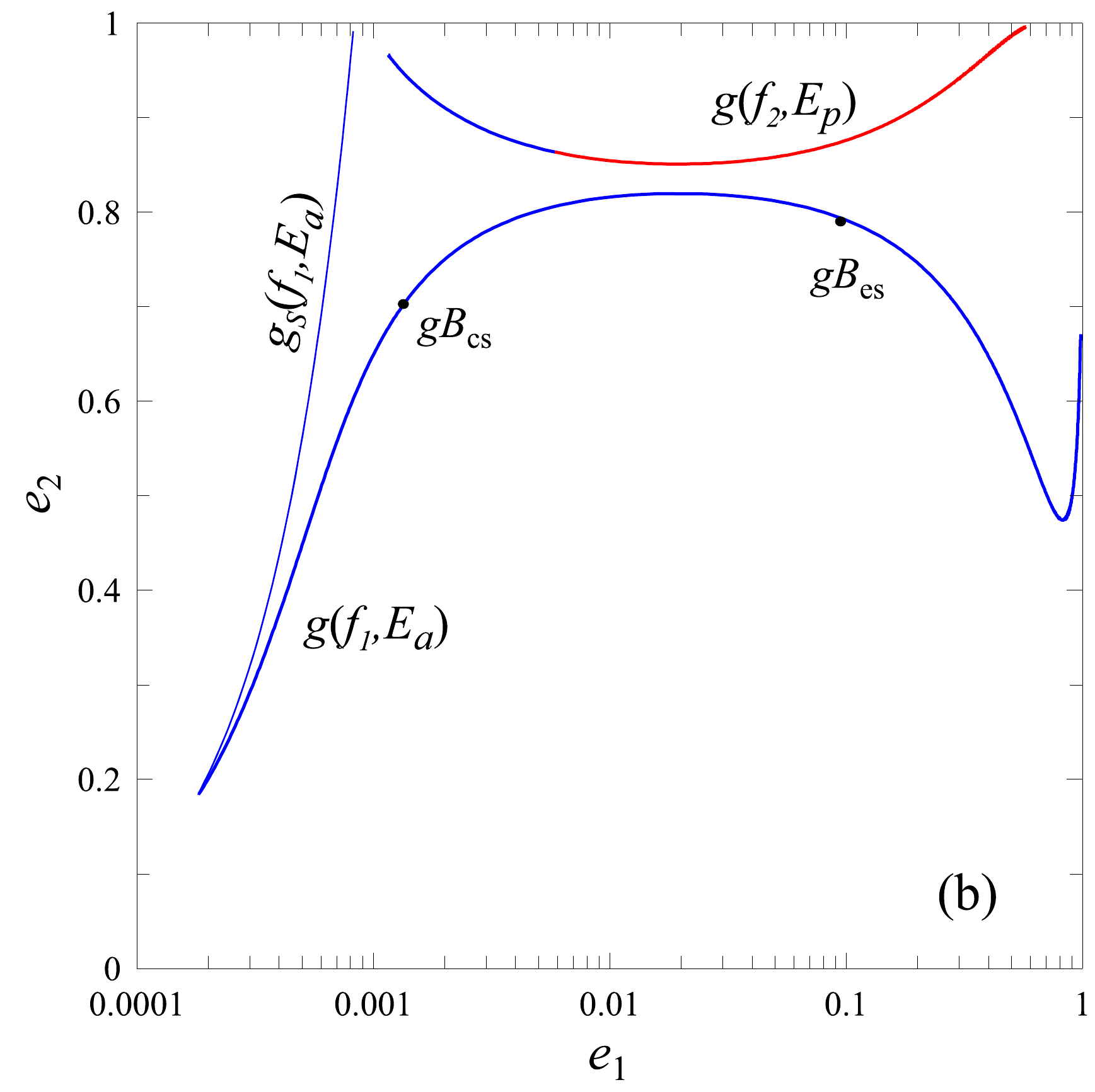}  
\end{array}$
\caption{Characteristic curves of the families $g(f_1,E_a)$ and $g(f_2,E_p)$ for $m_1=10^{-3}$ and $m_2=10^{-6}$ on the eccentricities plane. {\bf a} The characteristic curves close to the singular point $B_{ce}$. The families $f=f_1 \cup f_2$, $E_a$ and $E_p$ of the restricted problem are also presented. {\bf b} The total segments of the families (notice the logarithmic horizontal axis) including the segment $g_s(f_1,E_a)$ of satellite orbits. Blue (red) colour indicates horizontal stability (instability)}
\label{FIG_GP}
\end{figure}

In order to obtain the families formed in the P-GTBP, we firstly perform the continuation of an orbit of the family $E_a$ (or $E_p$) with respect to $m_2$ and we get a periodic orbit for a particular value $m_2\neq 0$. Then, by keeping fixed all the masses we perform continuation in the P-GTBP by using as parameter the variable $x_1$. Following this procedure, we obtain the families $g(f_1,E_a)$ and $g(f_2,E_p)$. The characteristic curves of the families in the eccentricity plane are shown in Fig. \ref{FIG_GP} for $m_1=10^{-3}$ and $m_2=10^{-6}$. In panel (a), a part of them near the singular point $B_{ce}$ is presented beside the families of the planar RTBP. The family $f$ of the PC-RTBP is presented by the line $e_1=0$ and is separated into two segments, $f_1$ and $f_2$, by the orbit $B_{ce}$. The transition of the characteristic curves from the restricted to the general problem has been discussed first by \citet{bozishadj76} and has been found in other resonances,  1:2 \citep{vkh09} and 1:3 and 3:2 \citep{avk11}. Particularly, as $m_2\neq 0$, the family $E_a$ and the family segment $f_1$ join smoothly forming the family $g(f_1,E_a)$ and the family $E_p$ joins the family segment $f_2$ forming the family $g(f_2,E_p)$. At the neighbourhood of the singular point $B_{ce}$, we obtain a gap between the two generated families. The formation of the two distinct families at this eccentricity domain causes a change in the topological structures in phase space, which may be related to that obtained in \cite{Leleu17}. The v.c.o. $B_{cs}$ and $B_{es}$ of the restricted problem, are continued for $m_2\neq 0$ as the orbits $gB_{cs}$ and $gB_{es}$, respectively. 

In Fig. \ref{FIG_GP}b, we present the complete characteristic curves of the families in the P-GTBP. We use the scale $log(e_1)$, in order to emphasize the structure for $e_1\approx 0$. All orbits of $g(f_1,E_a)$ are horizontally stable. They are also vertically stable except those in the segment between the v.c.o. $gB_{cs}$ and $gB_{es}$, which are vertically unstable. The continuation to the general problem does not affect the horizontal and vertical stability for sufficiently small values of $m_2$. The orbits of the family $g(f_2,E_p)$ which are continued from $f_2$ are horizontally stable, while its segment that originates from the $E_p$ family consists of unstable orbits. All orbits of $g(f_2,E_p)$ are vertically unstable. We mention also that for $g(f_1,E_a)$ it is $\theta=0$ and $\Delta \varpi=\pi$, while for $g(f_2,E_p)$ it is $\theta=\Delta \varpi=0$. 

In Fig. \ref{FIG_GP}b, we also present the family segment $g_s(f_1,E_a)$ of satellite periodic orbits which continues the family $g(f_1,E_a)$. In the eccentricity plane we obtain a cusp, where these families meet, which can be assumed as a border between planetary type orbits (like in $\textnormal{QS}_h$ domain) and satellite orbits (like in sRS domain) \citep{hpv09,hv11b}. However, in the variables of the rotating frame the two families join smoothly.  Along the family $g_s(f_1,E_a)$, the eccentricity $e_2$ seems to increase rapidly and takes values $>1$. This is due to its computation in the heliocentric frame. In the planetary frame, both $e_1$ and $e_2$ tend to zero. All orbits of $g_s(f_1,E_a)$ are both horizontally and vertically stable, in consistency with the stability of the family segment of $f$, where they originate from.                     

\subsection{The spatial general problem (S-GTBP)}
\begin{figure}[ht]
\centering
\includegraphics[width=0.7 \textwidth]{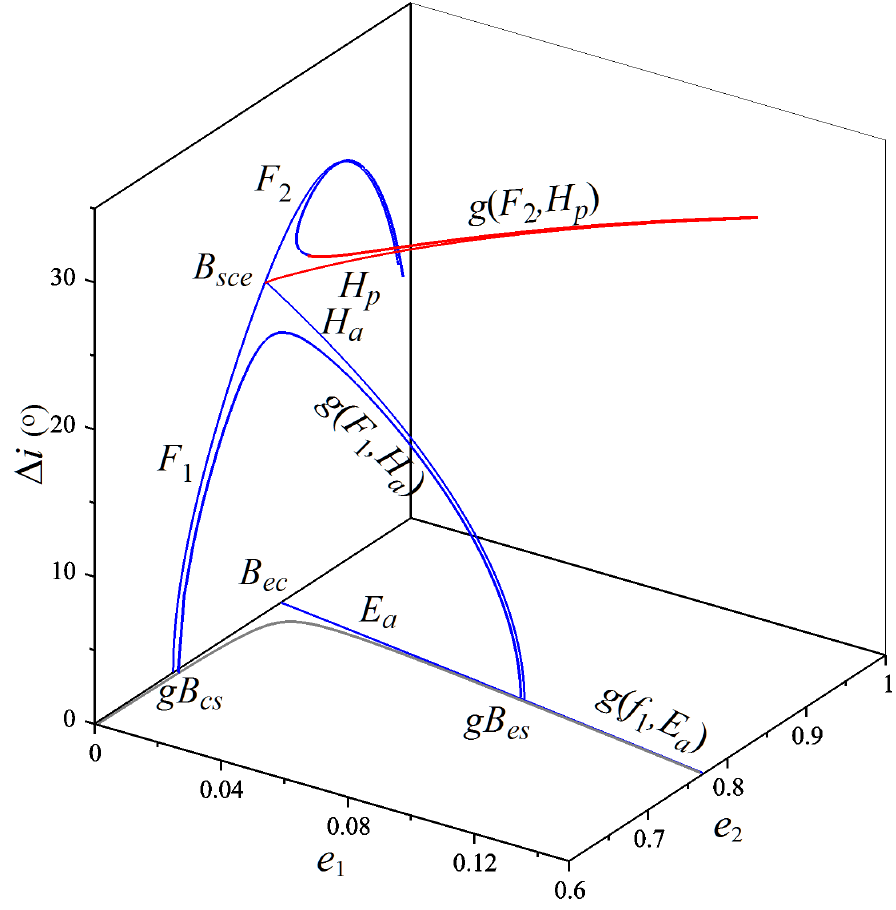}
\caption{The families $g(F_1,H_a)$ and $g(F_2,H_p)$ of the S-GTBP in the projection space $e_1-e_2-\Delta i$. Blue (red) color indicates linear stability (instability). The families of the restricted problems and the planar family $g(f_1,E_a)$ are also shown (family $f$ extends along the axis $e_1=0$, $\Delta i =0^\circ$)} 
\label{FIG_SGfams}
\end{figure}

Similarly to the planar problem, all orbits of the SC-RTBP are continued to the S-GTBP if their period is not an integer multiple of the period of the primaries \citep{kat79}. Also, the periodic orbits of the SE-RTBP ($e_1\neq 0$) are generically continued to the S-GTBP \citep{ichkm78}. Subsequently, all orbits of family $F$ are continued for $m_2\neq 0$ except the orbit $B_{sce}$, which has a period equal to the period of primaries ($T=2\pi$). This critical orbit separates the spatial family $F$ into two segments, $F_1$ and $F_2$ (Fig. \ref{FIG_SGfams}) and generates the families $H_p$ and $H_a$ of the SE-RTBP (see also Fig. \ref{FIG_Hfams}), which are also continued in the S-GTBP. 

Our computations for $m_1=10^{-3}$ and $m_2=10^{-6}$ show that a similar structure with that of the planar case is formed (Fig. \ref{FIG_SGfams}). Particularly, the continuation of the stable segment $F_1$ and the stable family $H_a$ constructs the stable family $g(F_1,H_a)$, which forms a bridge between the two orbits, $gB_{cs}$ and $gB_{es}$ of the planar family $g(f_1,E_a)$. Along the formed spatial family the orbits are symmetric with respect to the $Oxz$-plane and the initial conditions correspond to
$$
\theta=0, \;\;  \Delta \varpi=\pi, \;\; \Delta \Omega =\pi.
$$
The peak of the bridge corresponds to $e_1=0.0095$, $e_2=0.803$ and the maximum mutual inclination $\Delta i=20.1^\circ$. 

The continuation of the stable segment $F_2$ and the unstable family $H_p$ constructs for $m_2\neq 0$ the family $g(F_2,H_p)$. The stability changes close to the critical orbit $B_{sce}$, where there exists a gap between the two families. Along $g(F_2,H_p)$ the initial conditions correspond to           
$$
\theta=0, \;\;  \Delta \varpi=0, \;\; \Delta \Omega =0.
$$   

Computations of the bridge family $g(F_1,H_a)$ are also found in \citet{av14c}, but without explaining its origin. Also, in that paper, computations were performed by starting from the v.c.o. of the planar families given in \citet{hv11b}. However, by using such an approach, we were not able to detect the existence of the family $g(F_2,H_p)$.  

\section{Mass dependence of QS periodic motion}\label{var}

In Sect. \ref{res}, we presented the families of periodic orbits of the RTBP for $\mu=0.001$. Particularly, the structure of the families is described by the critical orbits $B_{cs}$,  $B_{ce}$, $B_{es}$ and $B_{sce}$ (see Fig. \ref{FIG_Hfams}). By performing numerical computations in the range $m_\oplus\leq \mu \leq 10m_J$ we found that there are no structural changes, namely all critical orbits and the corresponding families still exist and their stability, either horizontal or vertical, is unaltered. Therefore, the picture depicted in Fig. \ref{FIG_Hfams} holds for all values of $\mu$ in the above interval at least. The location of the critical orbits for some values of $\mu$ is given in Table \ref{tab1}.  
\begin{table}[ht]
\caption{Location of the critical orbits of the restricted model for some values of the mass parameter $\mu$} 
{
\begin{tabular}{|c|c|c|cc|cc|}
\hline 
   & $B_{cs}$ & $B_{ce}$ & \multicolumn{2}{|c|}{$B_{es}$} & \multicolumn{2}{|c|}{$B_{sce}$} \\
$\mu$ & $e_2$ & $e_2$ & $e_1$ & $e_2$ & $e_2$ & $i_2$ \\
\hline
$3\cdot 10^{-6}$ & $0.697$ & $0.835$ & $0.087$ & $0.798$ & $0.816$ & $22.6^\circ$ \\
$3\cdot 10^{-4}$ & $0.698$ & $0.836$ & $0.087$ & $0.799$ & $0.816$ & $22.6^\circ$ \\
$0.001$ & $0.700$ & $0.836$ & $0.086$ & $0.799$ & $0.816$ & $22.5^\circ$ \\
$0.004$ & $0.710$ & $0.837$ & $0.083$ & $0.801$ & $0.818$ & $22.2^\circ$ \\
$0.010$ & $0.723$ & $0.838$ & $0.077$ & $0.806$ & $0.821$ & $21.4^\circ$ \\
\hline
\end{tabular}
}
\label{tab1}
\end{table}  

\begin{figure}[ht]
\centering
\includegraphics[width=0.7 \textwidth]{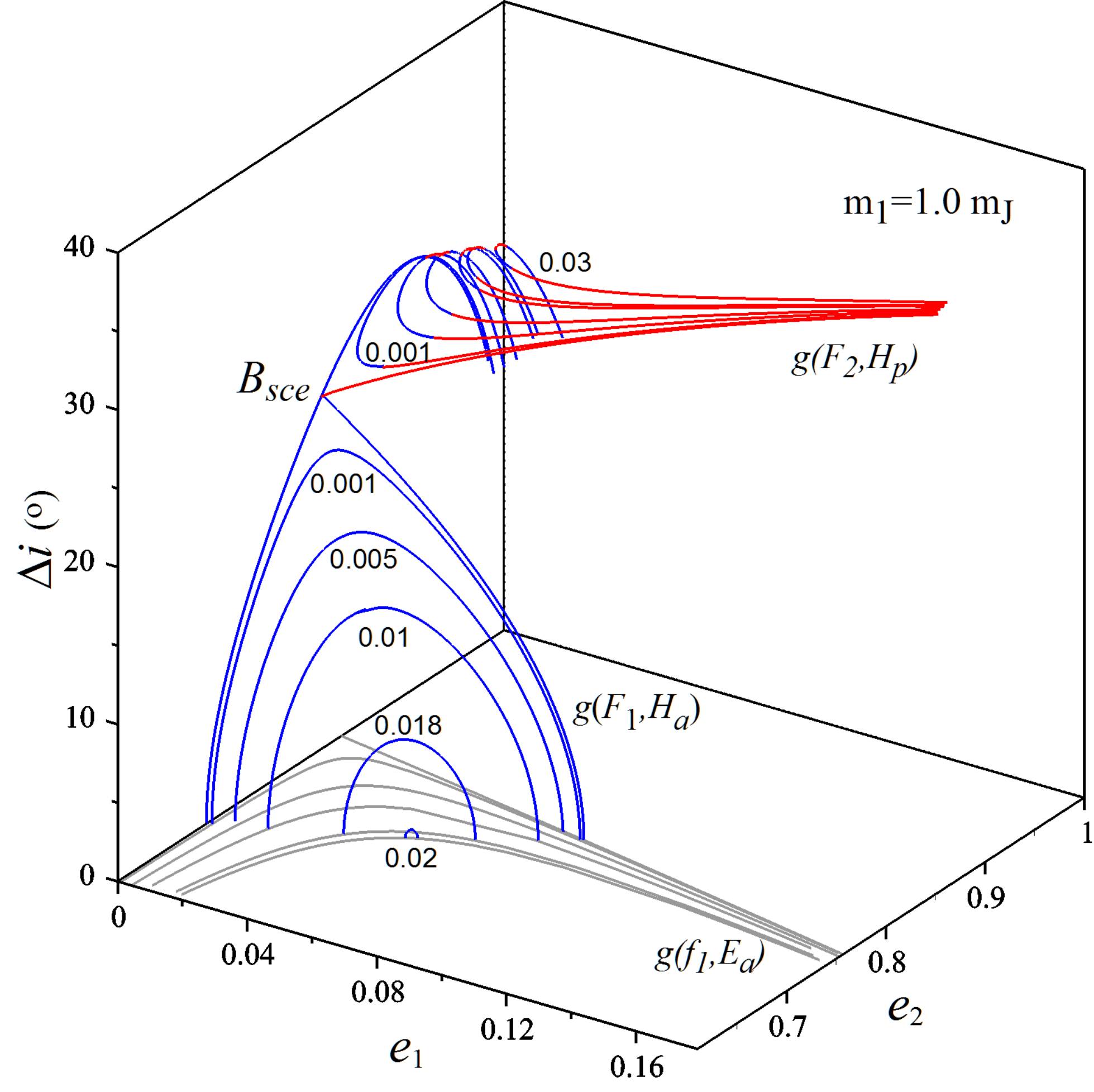}
\caption{The families $g(F_1,H_a)$ and $g(F_2,H_p)$ of the S-GTBP in the space $e_1-e_2-\Delta i$ for various mass-ratios $\rho=\frac{m_2}{m_1}$ mentioned in the labels and for fixed $m_1=0.001$. The presentation is similar as in Fig. \ref{FIG_SGfams}. The families $g(f_1,E_a)$ of the planar general problem are also shown in grey colour} 
\label{FIG_SGmj1}
\end{figure} 

In the GTBP, the two planetary masses are involved as parameters. For masses of the order of Jupiter and less, the location and the stability of families of the planar model do not seem to depend on the individual masses, but only on their ratio $\rho=m_2/m_1$ \citep{hpv09, hv11b}. Considering $m_1=0.001$, Fig. \ref{FIG_SGfams} shows the structure of families for $\rho=0.001$. For smaller mass ratios ($\rho \rightarrow 0$), we approach the picture of the RTBP. For larger values of $\rho$, we obtain that the critical orbits $gB_{cs}$ and $gB_{es}$ approach each other and coincide at a critical value $\rho*=0.0205$. Simultaneously, the ``bridge'' family $g(F_1,H_a)$ shrinks as $\rho$ increases and disappears at $\rho=\rho*$ (see Fig. \ref{FIG_SGmj1}). The linear stability of the family is unaltered. 

The family $g(F_2,H_p)$ is also continued as $\rho$ increases and its continuation is not restricted by the critical mass-ratio $\rho*$. It consists of two main segments, a stable and an unstable one, but for $\rho \gtrsim 0.005$ a small unstable segment appears inside the stable one. We note that in this case the numerical computation of the linear stability is quite ambiguous, since the linear stability appears very close to the critical case. We used long-term computations of the deviation vectors, as in \cite{gta18}, in order to conclude accordingly.    

By performing computations for $m_\oplus\leq \mu \leq 10m_J$ we obtain a similar structure for the families. The values of the critical mass ratio, $\rho*$, is shown in Table \ref{tab2}. Also in Table \ref{tab3}, we present the orbital elements for some representative orbits of the ``bridge'' family $g(F_1,H_a)$ for some values of $\rho$. 

\begin{table}[ht]
\caption{Critical mass-ratios $\rho ^*=\frac{m_2}{m_1}$ for various values of $m_1$} 
{\centering
\begin{tabular}{ll}
\hline \hline
$m_1$& $\rho ^*$  \\ 
\hline \hline
0.000003& 0.0212 \\
0.00001 & 0.0210 \\
0.0001  & 0.0209 \\
0.001   & 0.0205 \\
0.0015	& 0.0202 \\
0.002		& 0.0200 \\
0.005   & 0.0185 \\
0.01		& 0.0162 \\
\hline \hline
\end{tabular}
}
\label{tab2}
\end{table}

\begin{table}[h]
\caption{Orbital elements of spatial periodic orbits samples of family $g(F_1,H_a)$. For all orbits it is $\theta=0^\circ$, $\Delta\omega=0^\circ$, $\Delta\Omega=180^\circ$} 
{
\begin{tabular}{lcccc}
\hline \hline
$\rho$& $a_1/a_2$ & $e_1$ &$e_2$ & $i_1 (^{\circ})$   \\ 
\hline \hline
0.001		& 0.99807216 & 0.0022 & 0.7587 & 15.23 \\
				& 0.99856609 & 0.0123 & 0.8051 & 20.00 \\
				& 0.99869836 & 0.0474 & 0.8042 & 15.10 \\
				\hline
0.01		& 0.99804105 & 0.0197 & 0.7435 & 10.00 \\
				& 0.99833646 & 0.0289 & 0.7665 & 12.61 \\
				& 0.99860354 & 0.0558 & 0.7825 & 10.07 \\
				\hline
0.018		& 0.99807499 & 0.0350 & 0.7361 & 3.51 \\
				& 0.99829821 & 0.0443 & 0.7518 & 5.86 \\
				& 0.99848330 & 0.0591 & 0.7633 & 3.47 \\
				\hline
\hline
\end{tabular}
}
\label{tab3}
\end{table}

\begin{figure}[ht]
$\begin{array}{cc}
\includegraphics[height=6cm]{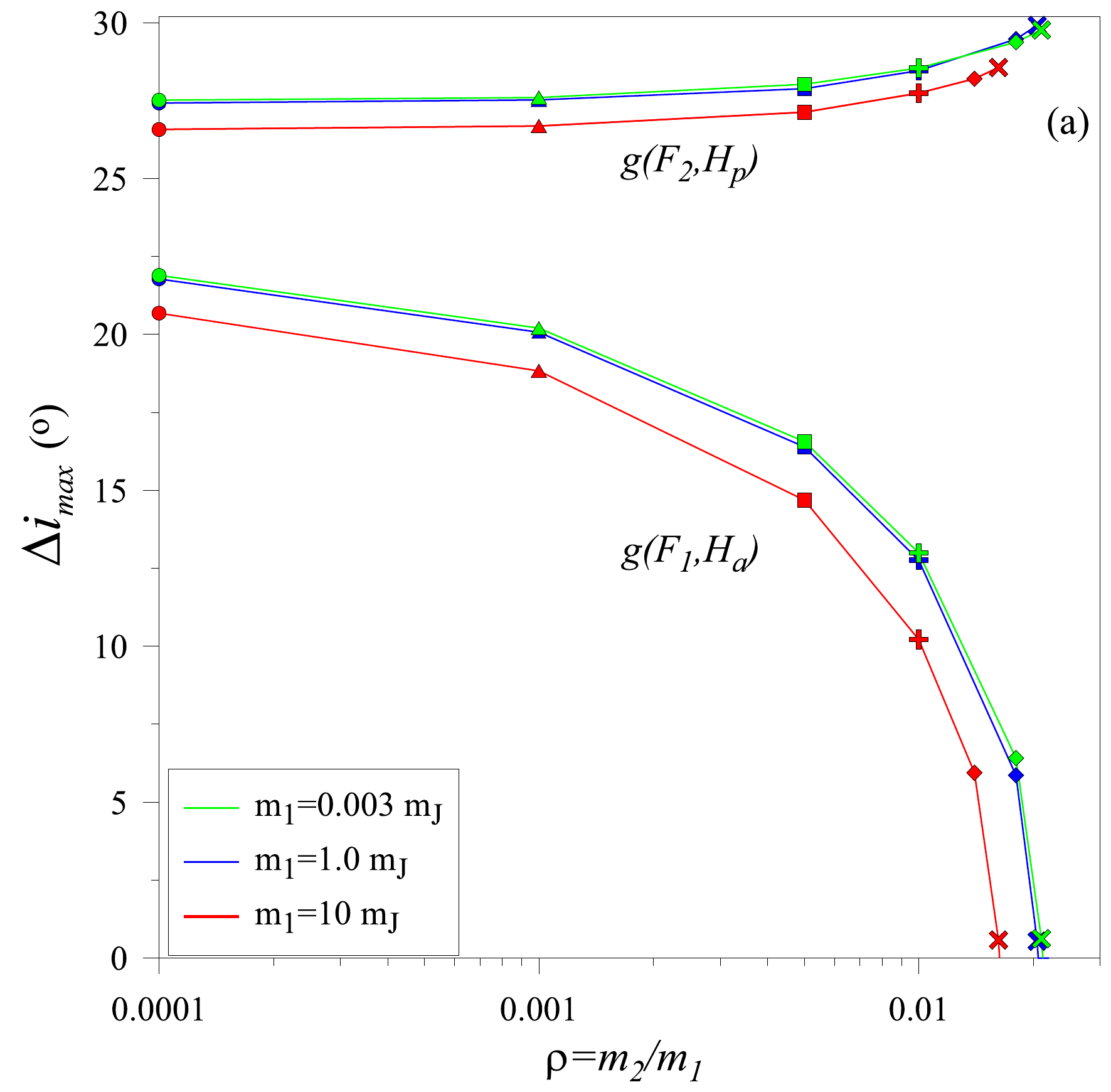}&\includegraphics[height=6cm]{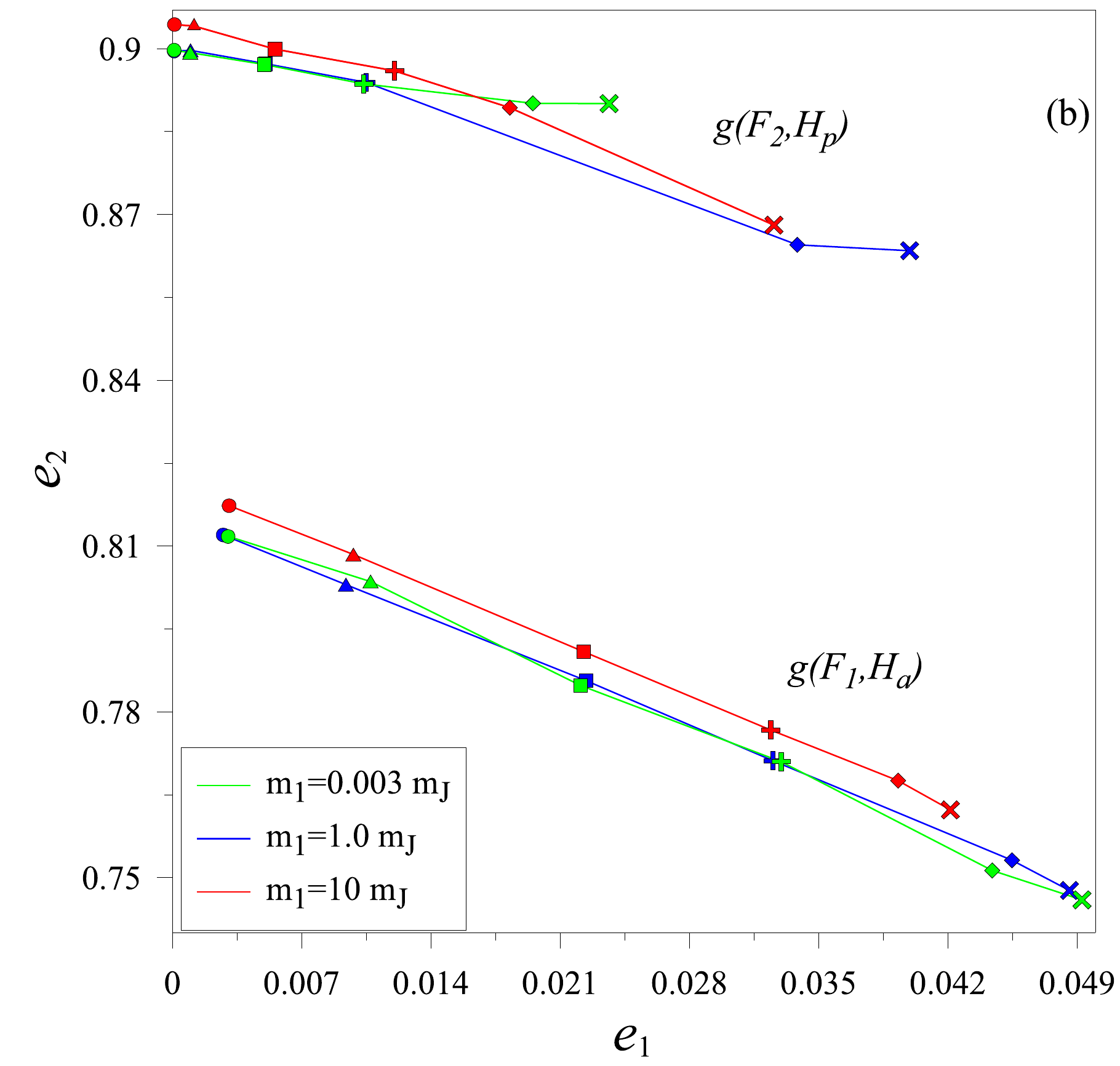}  \\
\end{array}$
\caption{{\bf a} The maximum mutual inclination, $\Delta i_{max}$, observed in families $g(F_1,H_a)$ and $g(F_2,H_p)$, when $m_1$ equals to $0.003 m_J$ (green), $1.0 m_J$ (blue) and $10 m_J$ (red) as the mass-ratio, $\rho$, varies. {\bf b} The eccentricity values of the orbits presented in panel {\bf a}} 
\label{FIG_dirho}
\end{figure}

The maximum mutual inclination observed along the families  $g(F_1,H_a)$ and  $g(F_2,H_p)$ is presented in the left panel of Fig. \ref{FIG_dirho} as a function of $\rho$. Along the ``bridge'', the maximum mutual inclination, $\Delta i\approx 22.5^\circ$, appears as $\rho\rightarrow 0$. For the family $g(F_2,H_p)$ the maximum $\Delta i$ increases as $\rho$ increases but the particular orbits seem to become unstable for $\rho \gtrsim 0.005$. In the right panel of Fig. \ref{FIG_dirho}, the  most mutually inclined orbits are presented on the eccentricity plane. It is clear that inclined periodic motion corresponds to low eccentricity value of the heavier planet (planet 1), but very high eccentricities for the lighter one (planet 2).    

\section{Conclusions}\label{con}
Our study concerns the quasi-satellite (QS) motion of the 1:1 mean-motion resonance which can find various applications in celestial mechanics.	We consider both the problems of QS motion of an asteroid in the framework of the RTBP and the QS planetary motion in the framework of the GTBP. We focus on the computation of families of periodic orbits by assuming the method of analytical continuation and applying differential corrections. It is well known that periodic orbits play an important role in the dynamics and their linear stability or instability indicates in general the existence of regions in phase space with stable or chaotic motion, respectively. 

For the case of asteroids (massless bodies), we started our study from the planar circular RTBP, where the backbone of QS motion is the horizontally stable family $f$. The horizontal and vertical stability of this family indicates also the existence of long-term stability, when considering small perturbations by adding a small eccentricity in the motion of the primaries or by assuming spatial orbits of small inclination. These results have been verified also by other studies \citep[e.g.][]{Mikkola06, Sidorenko14, Pousse17}. We determined two critical orbits along family $f$, called $B_{cs}$ and $B_{ce}$. The critical orbit $B_{cs}$ separates the family $f$ in two segments, one vertically stable and one vertically unstable. Therefore, $B_{cs}$ is a v.c.o., which is continued in the spatial model by adding inclination to the asteroid, and hence, we obtain the family $F$ of spatial periodic QS orbits. The $B_{ce}$ orbit belongs to the segment of vertically unstable orbits of $f$. It has period $T=2\pi$ and is continued in the elliptic model and generates two families of planar periodic orbits, $E_a$ and $E_p$. Both critical orbits and all orbits of the above mentioned generated families are highly eccentric orbits for the massless body. Also, along family $F$ the inclination reaches the value of $27^{\circ}$.

Family $E_a$ contains the critical orbit $B_{es}$ which is a v.c.o. Namely, at $B_{es}$ the family $E_a$ turns from vertically unstable to vertically stable and a new stable family, $H_a$ is derived by continuation in the spatial model. The spatial family $F$ contains a critical orbit $B_{cse}$, which has period $T=2\pi$ and is continued in the spatial elliptic model generating two new families. Our computations showed that one of the families, which arises from $B_{cse}$ coincides with $H_a$ (it is the same family), which emanates from the planar v.c.o. $B_{es}$. So, the family $H_a$ of the spatial elliptic RTBP forms a bridge between the families $E_a$ and $F$ of the elliptic planar and the spatial circular RTBP, respectively. The second generated family, $H_p$ is unstable and extends up to very high eccentricity values. This structure of periodic solutions in phase space holds at least in the mass range $m_\oplus\leq \mu \leq 10m_J$. 
  
Apart from the isolated critical orbits $B_{ce}$ and $B_{cse}$ all other orbits of the above mentioned planar and spatial families are continued by adding mass, $m_2$, to the massless body, i.e. by passing from the RTBP to GTBP. We showed that for $m_2\neq 0$ two families of inclined orbits are formed, called $g(F_1,H_a)$ and $g(F_2,H_p)$. Family $g(F_1,H_a)$ is stable and forms a bridge between two orbits of the QS family of the planar GTBP \citep{Giuppone10,hv11b} and reaches a maximum inclination value that depends mainly on the mass-ratio $\rho=m_2/m_1$. This ``bridge'' becomes lower and lower as $\rho$ increases and disappears for $\rho\approx 0.02$.  The family $g(F_2,H_p)$ is located at higher inclinations than those of the ``bridge'' and it consists mainly of two segments, one stable and one unstable. From a qualitative point of view, the above structure of periodic QS motion is almost unaltered for $m_\oplus\leq \mu \leq 10m_J$.             
 
The QS periodic orbits studied in this paper consist the exact 1:1 resonant solutions. Particularly, stable periodic solutions should form islands in phase space, where the resonant angle $\theta=\lambda_2-\lambda_1$ librates regularly. Considering a particular TBP model and a stable periodic orbit of it, we obtain in its vicinity librations for the resonant angle $\Delta \varpi=\varpi_2-\varpi_1$, too. For the planar models, such librations have been indicated by the studies cited along the paper. The existence of inclined librations close to spatial periodic orbits has been checked and verified by numerical integrations, though they are not presented in this paper. The width of the area of inclined librations, for both asteroid and planetary QS orbits, requires further studies. 
		
\vspace{1cm}
\noindent
{\bf Acknowledgements.} The work of KIA was supported by the University of Namur. 

\vspace{1cm}
\noindent
{\bf Conflict of Interest.} The authors declare that they have no conflict of interest. 

\bibliographystyle{plainnat}
\bibliography{gvnbib2} 
\end{document}